\documentclass[twoside,11pt]{article}

\usepackage[cp1251]{inputenc}

\usepackage{graphicx}

\usepackage{amssymb}
\usepackage{amsmath}
\usepackage{amsfonts}
\usepackage{wasysym}

\begin{document}           

\title{\Large 
Quantum corrections to the dynamics of the expanding universe
}
\author{V.E. Kuzmichev, V.V. Kuzmichev\\[0.5cm]
\itshape Bogolyubov Institute for Theoretical Physics,\\
\itshape National Academy of Sciences of Ukraine, Kiev, 03680 Ukraine}

\date{}

\maketitle

\begin{abstract}
The dynamics of the expanding universe is analyzed in terms of the quantum geometrodynamical model. It is shown that the equations of 
quantum theory in the form of the eigenvalues equation similar to the stationary Schr\"{o}dinger equation complemented 
by the equations of motion for the momentum operator and its time derivative in Heisenberg's form reduce to the Einstein equations
with an additional source of the gravitational field of quantum nature. The spatially closed universe with cosmological constant, originally filled with 
a uniform scalar field and radiation, is considered as quantum cosmological system. The perfect fluid in the form of radiation defines
the material reference frame. The properties of the averaged scalar field which acts like ordinary matter are investigated.
After averaging over its quantum states, the free scalar field turns into the Weyssenhoff fluid characterized
by the energy density, pressure, and spin of constituent particles. The cases when the contribution of the quantum effects into the gravitational 
interaction becomes significant on macroscopic scale are analyzed. It is demonstrated that, unless the whole, at least a part of such 
matter-energy constituents as dark matter and dark energy may have a quantum origin.
\end{abstract}


\section{Introduction}
According to the modern view, most of our universe consists of dark matter and dark energy,
whose nature is unknown \cite{Ben, Pla}. One cannot exclude that at least a part of these matter constituents, unless the whole matter-energy, has a quantum origin, being,
in this case, a demonstration of quantum principles which reveal themselves in the universe on macroscopic (cosmological) scales. Another
manifestation of quantum nature of the universe may be the existence of proper angular momenta (spins) 
of particles which compose matter in the universe. It is possible that the observed universe is a realization of a specific state of a more
general cosmological system described by the wave function which is a superposition of all admissible states satisfying a 
Schr\"{o}dinger-type equation. It can be considered as some implementation of the multiverse hypothesis (see, e.g., Refs. \cite{Dav,Teg}). In any case, the search for
macroscopical effects which give evidence concerning the quantum nature of our universe as a whole does make sense. For this purpose,
the appropriate quantum model of the universe is required. 

A consistent quantum theory of gravity, in principle, can be constructed on the basis of the Arnowitt-Deser-Misner
(ADM) Hamiltonian formalism \cite{ADM} of general relativity with the application of the 
canonical quantization method. The structure of constraints in general relativity is such that variables which correspond to true 
dynamical degrees of freedom cannot be singled out from canonical variables of geometrodynamics. This difficulty is stipulated by an absence of 
predetermined way to identify spacetime events in generally covariant theory \cite{Kuch}. The famous Wheeler-DeWitt equation 
\cite{Whe,DeW} of canonical quantum gravity does not refer to time variable. In principle, time can be reintroduced in the theory \cite{Ish}. 
The material reference frame allows to mark spacetime events. 
Perfects fluids are a special case of the relativistic elastic media with clocks proposed by DeWitt \cite{Bro}. 
They can be used to define the reference frame as a dynamical system.
In the framework of realistic description of the reference medium, the Dirac constraint quantization leads to the functional 
Schr\"{o}dinger-type equation with respect to a time variable which describes the evolution of 
the system from one spacelike hypersurface to another.
In a series of papers \cite{Kuz}, the quantum geometrodynamical approach with a 
well-defined time variable was developed for the description of the FRW universe.

In Sect.~2 we give the basic equations of quantum theory reduced to the form convenient for purposes of this paper.
In Sect.~3 the properties of the scalar field $\phi$ in the $\phi^{\alpha}$-model of interaction are studied.
The equations of motion for such a model are obtained in Sect.~4.
The properties of the averaged scalar field which acts like ordinary matter are investigated in Sect.~5.
Sect.~6 is devoted to the study of the influence of the quantum source of the gravitational field on the dynamics of the universe.
In Sect.~7 the exact solution of the non-linear equation for the phase of the wave function in the model of the free scalar field is found.
In Sect.~8 it is shown that the averaged free scalar field turns into the Weyssenhoff fluid.
In a short conclusion some obtained results are summarized.
Some equations in the ordinary physical units are given in Appendix for illustration.

\section{Basic equations}
The universe which is a homogeneous, isotropic, and spatially closed quantum system is considered. We assume that such a universe
is originally filled with a uniform scalar field  $\phi$ and a perfect fluid. 
After averaging with respect to appropriate quantum states, the scalar field turns into the effective barotropic fluid 
to which we shall refer as the $\phi$-substance.
The perfect fluid defines a material reference frame \cite{Bro, Kuz} and it is taken in the form of relativistic matter (radiation) with the energy density $\rho_{\gamma} = E / a^{4}$, where 
$E = \mbox{const}$, $a$ is the cosmic scale factor, and the equation of state $p_{\gamma} = \frac{1}{3} \rho_{\gamma}$, $p_{\gamma}$
is the pressure of radiation. The stationary states of such a quantum system with a definite value of $E$ are described by the wave function of 
two variables $\psi (a, \phi) = \langle a, \phi | \psi \rangle$ which satisfies the eigenvalues equation of Schr\"{o}dinger-type. If the density 
$\rho_{\gamma}$ and the scale factor $a$ are measured in GeV/cm$^{3}$ and cm, respectively, the eigenvalue $E$ has the dimensions GeV 
cm. It is convenient to pass to the dimensionless quantities using the modified Planck system of units. We use the $l_{P} = \sqrt{2 G \hbar / (3 \pi 
c^{3})}$ as a unit of length and the $\rho_{P} = 3 c^{4} / (8 \pi G l_{P}^{2})$ as a unit of energy density and pressure. The proper time $\tau$ is
taken in units of $t_{P} = l_{P} / c$ and the scalar field is measured in $\phi_{P} = \sqrt{3 c^{4} / (8 \pi G)}$. The mass-energy is taken in 
units of Planck mass $m_{P} c^{2} = \hbar c / l_{P}$. Throughout the paper, the equations are given for dimensionless quantities, 
unless otherwise stipulated.

Let us briefly address the introduction of time variable in our approach.
The universe under consideration is described by the Robertson-Walker metric
\begin{equation}\label{01}
     ds^{2} = a^{2}(\eta) [N^{2}(\eta) d\eta^{2} - d\Omega_{3}^{2}],
\end{equation}
where $N$ is the lapse function that specifies the time reference scale, $d\Omega_{3}^{2}$ is a 
line element on a unit three-sphere. Using the ADM formalism, the action in the case under study is reduced to the form
(see Refs. \cite{Kuz} for details)
\begin{equation}\label{03}
    I = \int d\eta \left\{\pi_{a}\,\frac{da}{d\eta} + \pi_{\phi}\,\frac{d\phi}{d\eta} + 
         \pi_{\Theta}\,\frac{d\Theta}{d\eta} + 
\pi_{\tilde{\lambda}}\,\frac{d\tilde{\lambda}}{d\eta} - H \right\},
\end{equation}
where  $\pi_{a},\, \pi_{\phi},\, \pi_{\Theta},\, \pi_{\tilde{\lambda}}$ are the momenta 
canonically conjugate with the variables $a,\, \phi,\, \Theta,\, \tilde{\lambda}$,
\begin{eqnarray}\label{04}
    H & = & \frac{N}{2} \left\{-\,\pi_{a}^{2} - a^{2} + a^{4} [\rho_{\phi} + \rho_{\gamma} + \rho_{\Lambda}]\right\} \nonumber \\ 
& + & \lambda_{1}\left\{\pi_{\Theta} - \frac{1}{2}\,a^{3} \rho_{0} s\right\}
+ \lambda_{2}\left\{\pi_{\tilde{\lambda}} + \frac{1}{2}\,a^{3} \rho_{0} \right\},
\end{eqnarray}    
is the Hamiltonian, 
\begin{equation}\label{05}
    \rho_{\phi} = \frac{2}{a^{6}}\,\pi_{\phi}^{2} + V(\phi)
\end{equation}
is  the energy density of a scalar field $\phi$ with the potential $V(\phi)$, 
$\rho_{\gamma} = \rho_{\gamma}(\rho_{0}, s)$ is the energy density of a perfect fluid\footnote{At this point an arbitrary perfect fluid
is assumed. The subscript $\gamma$ is used having in mind the future choice of a perfect fluid in the form of radiation.}
which is a function of the density of the rest mass $\rho_{0}$ and the specific entropy $s$,
$\rho_{\Lambda}$ is the vacuum energy density of the field $\phi$ with the equation of state $p_{\Lambda} = - \rho_{\Lambda}$,
$\rho_{\Lambda} = \frac{\Lambda}{3}$, $\Lambda$ is a cosmological constant, $p_{\Lambda}$ is the pressure.  
The $\Theta$ is the thermasy (potential for the temperature, $\mathcal{T} = \Theta_{,\,\nu} 
U^{\nu}$). The $\tilde{\lambda}$ is 
the potential for the specific free energy $f$ taken with an inverse sign, $f = 
-\,\tilde{\lambda}_{,\,\nu} U^{\nu}$. The $U^{\nu}$ is the four-velocity.
The momenta $\pi_{\rho_{0}}$ and $\pi_{s}$
conjugate with the variables $\rho_{0}$ and $s$ vanish identically,
\begin{equation}\label{06}
    \pi_{\rho_{0}} = 0, \qquad \pi_{s} = 0.
\end{equation}
The Hamiltonian (\ref{04}) of such a system has the form of a linear 
combination of constraints and weakly vanishes,
\begin{equation}\label{07}
    H \approx 0,
\end{equation}
where the sign $\approx$ means that Poisson brackets must all be worked out 
before the use of the constraint equations.
The $N$, $\lambda_{1}$, and $\lambda_{2}$ are Lagrange multipliers.
The variation of the action (\ref{03}) with respect to them leads to three constraint 
equations
\begin{equation}\label{08}
    -\,\pi_{a}^{2} - a^{2} + a^{4} [\rho_{\phi} + \rho_{\gamma} + \rho_{\Lambda}] \approx 0, \quad
    \pi_{\Theta} - \frac{1}{2}\,a^{3} \rho_{0} s \approx 0, \quad
     \pi_{\tilde{\lambda}} + \frac{1}{2}\,a^{3} \rho_{0} \approx 0.
\end{equation}

From the conservation of these constraints in time it follows that
the conservation laws hold,
\begin{equation}\label{09}
    E_{0} \equiv a^{3} \rho_{0} = \mbox{const}, \qquad s =  \mbox{const},
\end{equation}
where the first relation describes the conservation law of a
macroscopic value which characterizes the number of particles of a perfect fluid, the 
second equation represents the conservation of the specific entropy.
Taking into account these conservation laws 
and the equations (\ref{06}) one can discard degrees of freedom corresponding to the 
variables $\rho_{0}$ and $s$, and convert the second-class constraints into
first-class constraints \cite{Kuz} in accordance with Dirac's proposal.

The equation of motion for the classical dynamical variable $\mathcal{O} = 
\mathcal{O}(a, \phi, \pi_{a}, \pi_{\phi}, \dots )$ has the form
\begin{equation}\label{010}
    \frac{d\mathcal{O}}{d\eta} \approx \{\mathcal{O}, H\},
\end{equation}
where $H$ is the Hamiltonian (\ref{04}), $\{.,.\}$ are Poisson brackets. 

In quantum theory first-class constraint equations (\ref{08}) become constraints on the
state vector $\Psi$. Passing from classical variables to corresponding operators,  
using the conservation laws (\ref{09}), and introducing the non-coordinate co-frame
\begin{eqnarray}\label{012}
  h\, d\tau = s\,d\Theta\, -\,d\widetilde{\lambda},\qquad
  h\, dy = s\,d\Theta\, +\,d\tilde{\lambda},
\end{eqnarray}
where $h = \frac{\rho_{\gamma} + p_{\gamma}}{\rho_{0}}$ is the specific enthalpy which plays the role
of inertial mass, $p_{\gamma}$ is the pressure of a perfect fluid, $\tau$ is proper time in every 
point of space, and $y$ is supplementary variable\footnote{The corresponding derivatives commute 
between themselves, $\left[\partial_{\tau},\,\partial_{y}\right] = 0$.},
we obtain three equations \cite{Kuz}
\begin{equation}\label{014}
    \left\{-\, i\,\partial_{\tau_{c}} - \frac{1}{2}\, E_{0} \right\} \Psi = 0,
 \qquad \partial_{y} \Psi = 0,
\end{equation}
\begin{equation}\label{015}
    \left\{-\,\partial^{2}_{a} +  a^{2} - 2 a \hat{H}_{\phi} - a^{4} \rho_{\Lambda} - E\right\} \Psi = 0,
\end{equation}
where  $\tau_{c}$ is the time variable connected with the proper time $\tau$ by the differential relation  $d\tau_{c} = h\, d\tau$,
and a perfect fluid is chosen in the form of relativistic matter. The operator
\begin{equation}\label{2}
\hat{H}_{\phi} = \frac{1}{2} a^{3} \hat{\rho}_{\phi} \quad \mbox{with} \quad 
\hat{\rho}_{\phi} = - \frac{2}{a^{6}} \partial_{\phi}^{2} + V(\phi)
\end{equation}
plays the role of a Hamiltonian of the scalar field with the operator of energy density $\hat{\rho}_{\phi}$, the quantity $a^{3} / 2$ is the proper volume. From the equations (\ref{014}) it follows that $\Psi$ does not depend on the variable $y$. 
The first equation of the set (\ref{014}) has a particular solution in the form
\begin{equation}\label{017}
    \Psi = \mbox{e}^{\,\frac{i}{2}\,E (T - T_{0})} |\psi (T_{0}) \rangle,
\end{equation}
where $T$ is the rescaled time variable, $dT= \frac{E_{0}}{E} d \tau_{c} = N d\eta$.
The state vector $|\psi\rangle$ is defined in the space of two variables  $a$ and $\phi$,
and determined by the stationary Schr\"{o}dinger-type equation
\begin{equation}\label{1}
\left( - \partial_{a}^{2} + a^{2} - 2 a \hat{H}_{\phi} - a^{4} \rho_{\Lambda}\right) | \psi \rangle = E | \psi \rangle.
\end{equation}
The vector $|\psi \rangle$ represents the dynamical state of the universe at some instant of time $T_{0}$.
For the universe without radiation, such a procedure leads to the Wheeler-DeWitt equation of a `minisuperspace model' 
which has a form of Eq. (\ref{1}) with $E = 0$.

Considering the vector $|\psi \rangle$ as immovable vector of the Heisenberg representation, we have the following equation of motion
\begin{equation}\label{019}
      \langle \psi|\frac{1}{N}\, \frac{d}{d\eta} \hat{\mathcal{O}}|\psi\rangle =
    \frac{1}{N}\, \frac{d}{d\eta}\langle \psi|\hat{\mathcal{O}}|\psi\rangle =
     \frac{1}{i} \, \langle \psi|[\hat{\mathcal{O}},\frac{1}{N} \hat{H}]|\psi\rangle, 
\end{equation}
where $[.,.]$ is a commutator, and $\hat{H}$ is determined by the expression
(\ref{04}), in which all dynamical variables are substituted with operators. The 
observable $\hat{\mathcal{O}}$  corresponds to the classical dynamical variable 
$\mathcal{O}$. Let  $\hat{\mathcal{O}} = a$, then
\begin{equation}\label{3}
\langle \psi |- i  \partial_{a}| \psi \rangle = \langle \psi |- \frac{da}{dT}| \psi \rangle.
\end{equation}
If $\hat{\mathcal{O}} = - i\partial_{a}$, then
\begin{equation}\label{4}
\langle \psi |- i  \frac{d}{dT} \partial_{a}| \psi \rangle = \langle \psi |a - \hat{H}_{\phi} + 3 \hat{L}_{\phi} - 2 a^{3} \rho_{\Lambda}| \psi \rangle,
\end{equation}
where the operator $\hat{H}_{\phi}$ is a Hamiltonian (\ref{2}) and
\begin{equation}\label{5}
\hat{L}_{\phi} = \frac{1}{2} a^{3} \hat{p}_{\phi} \quad \mbox{with} \quad 
\hat{p}_{\phi} = - \frac{2}{a^{6}} \partial_{\phi}^{2} - V(\phi)
\end{equation}
can be interpreted as a Lagrangian of the scalar field, $\hat{p}_{\phi}$ is
the operator of pressure.

\section{The $\phi^{\alpha}$-model}
The Hamiltonian $\hat{H}_{\phi}$ can be diagonalized by means of the state vectors $\langle x|u_{k} \rangle$, where 
$k$ is an index of the state, in the representation of some generalized variable $x = x(a, \phi)$. Assuming that the states $|u_{k} \rangle$
are orthonormalized, $\langle u_{k}|u_{k'} \rangle = \delta_{k k'}$, we obtain
\begin{equation}\label{6}
\langle u_{k}| \hat{H}_{\phi} |u_{k'} \rangle = M_{k} (a) \delta_{k k'},
\end{equation}
where the index $k$ can take both discrete and continuous values depending on the form of the potential $V(\phi)$. In general case, the value
$M_{k} (a)$ depends on $a$ and describes a classical source (as a mass-energy) of the gravitational field in $k$-th state. Its explicit form is
determined by the interaction model of the scalar field.

Let the potential $V(\phi)$ has a form
\begin{equation}\label{7}
V(\phi) = \lambda_{\alpha} \phi^{\alpha},
\end{equation}
where $\lambda_{\alpha} = \mbox{const}$, $\alpha$ is an arbitrary non-negative value, $\alpha \geq 0$. 

We make a scale transformation of the field $\phi$ and introduce a variable (a new field)
\begin{equation}\label{8}
x = \left(\frac{\lambda_{\alpha} a^{6}}{2} \right)^{\frac{1}{2 + \alpha}} \phi,
\end{equation}
which changes in the interval $- \infty < x < + \infty$ like the field $\phi$. In new variables $a$ and $x$, the Hamiltonian
$\hat{H}_{\phi}$ takes the form
\begin{equation}\label{9}
\hat{H}_{\phi} =  \left(\frac{\lambda_{\alpha}}{2} \right)^{\frac{2}{2 + \alpha}} a^{\frac{3 (2 - \alpha)}{2 + \alpha}} 
\left[-\partial_{x}^{2} + x^{\alpha} \right].
\end{equation}
The expression for the Lagrangian $\hat{L}_{\phi}$ (\ref{5}) is obtained from Eq. (\ref{9}) after the substitution 
$x^{\alpha} \rightarrow - x^{\alpha}$.

We introduce the state vectors $\langle x|u_{k} \rangle$ which satisfy the equation
\begin{equation}\label{10}
\left(-\partial_{x}^{2} + x^{\alpha} - \epsilon_{k} \right) |u_{k} \rangle = 0,
\end{equation}
where $\epsilon_{k}$ is eigenvalue. For $\alpha = 0$, this equation is the equation of free motion of the quantum analogue particle with
unit mass and doubled kinetic energy $\epsilon_{k} = k^{2} + 1$, where $k \geq 0$. For $\alpha = 1$, Eq. (\ref{10}) is the equation 
for Airy function. For $\alpha = 2$, it describes the quantum oscillator. For $\alpha = 4$, the asymptotics of the solution of Eq. (\ref{10}) is
expressed through the cylindrical function, $|u_{k} \rangle \sim \sqrt{x} Z_{1/6} (i x^{3} / 3)$, where 
$Z_{1/6} = c_{1} J_{1/6} + c_{2} N_{1/6}$, $J_{\nu}$ and $N_{\nu}$ are the Bessel functions of the first and second kind for $\nu = 1/6$, respectively \cite{Kam}.
Using the explicit form of the asymptotics, one can find the spectrum of values $\epsilon_{k}$ by numerical integration of Eq. (\ref{10}) 
with $\alpha = 4$.

The $\phi^{\infty}$-model with extremely strong self-action of the field $\phi$ which may occur in the very early universe (on sub-Planck
scales) is of peculiar interest.
Such a matter has only quantum properties. But its eigenstates $|u_{k} \rangle$ and eigenvalues $\epsilon_{k}$ cannot be find
directly from Eq. (\ref{10}), since for $\alpha = \infty$ it is senseless. This case requires a separate consideration.

We show that the problem here reduces to the equation which describes the motion of the analogue particle in the infinitely deep potential well
with zero value of the potential on the interval $- \varepsilon < x < \varepsilon$, where $\varepsilon > 0$, and $\varepsilon \ll 1$.
The potential well is bounded by an infinitely high potential barrier for $|x| > \varepsilon$. Really, the potential (\ref{7})
at $\alpha = \infty$ is equal to $V(\phi) = \lambda_{\infty} \phi^{\infty} \equiv \frac{1}{2} V_{0}$. It becomes infinite for finite values of
$\lambda_{\infty}$. But in the region, where the coupling constant $\lambda_{\infty} = 0$, the quantity $V_{0}$ may have a
finite value, including zero. The Hamiltonian (\ref{2}) takes the form
\begin{equation}\label{11}
\hat{H}_{\phi} = \frac{a^{3}}{4} \left(-\partial_{x}^{2} + V_{0} \right),
\end{equation}
where $x = \frac{a^{6}}{4} \phi$, and Eq. (\ref{10}) is substituted by the equation
\begin{equation}\label{12}
\left(-\partial_{x}^{2} + V_{0} - \epsilon_{k} \right) |u_{k} \rangle = 0.
\end{equation}
Setting $V_{0} = 0$ in the domain $|x| < \varepsilon$, where $\lambda_{\infty} = 0$, and 
$V_{0} = \infty$ for $|x| > \varepsilon$, where $\lambda_{\infty} \neq 0$, one can write the general solution of Eq. (\ref{12})
in the form of linear combination
\begin{equation}\label{13}
\langle x|u_{k} \rangle = A_{k} \cos \left(\frac{k \pi}{2 \varepsilon}x \right) \delta_{k, 2n + 1} + 
B_{k} \sin \left(\frac{k \pi}{2 \varepsilon}x \right) \delta_{k, 2n},
\end{equation}
where $n = 0,1,2, \dots$, $A_{k}$ and $B_{k}$ are normalization constants for even and odd states with respect to inversion
$x \rightarrow -x$ (see, e.g. Refs. \cite{Sch, Mes}). This solution corresponds to the discrete spectrum 
\begin{equation}\label{14}
\epsilon_{k} = \left(\frac{k \pi}{2 \varepsilon} \right)^{2}, \quad k = 0,1,2, \dots, \infty.
\end{equation}
Passing in the argument of the function $|u_{k} \rangle$ to the variable $\phi$ and setting it equal to $\frac{k \pi}{2} \phi$,
we obtain the restriction on the range of values of the scale factor, where $\lambda_{\infty} = 0$, and $\varepsilon \ll 1$:
$a \ll 2^{1/3}$. Since, in accepted dimensionless units, the value $a = 1$ corresponds to the Planck length $l_{P}$,
the domain with vanishing coupling constant $\lambda_{\infty}$ may be realized on sub-Planck scales only.

\section{The equation of motion in the $\phi^{\alpha}$-model}
Let us consider the consequences in the equations of motion (\ref{1}), (\ref{3}), and (\ref{4}), to which 
the model of the scalar field with the potential (\ref{7}) leads. Assuming that the states $|u_{k} \rangle$ are
orthonormalized, from Eqs. (\ref{6}), (\ref{9}), and (\ref{10}), we obtain the expression for classical source of the gravitational field
\begin{equation}\label{15}
M_{k} (a) =  \epsilon_{k} \left(\frac{\lambda_{\alpha}}{2} \right)^{\frac{2}{2 + \alpha}} a^{\frac{3 (2 - \alpha)}{2 + \alpha}}
\end{equation}
(here and below, we do not indicate the explicit dependence of the calculated quantities on $\alpha$, assuming that it exists).

We look for the solution of Eq. (\ref{1}) in the form of a superposition of the states with different $M_{k} (a)$. Within the framework of
the multiverse concept, it could mean that the wave function of the system $|\psi \rangle$ is considered as a superposition of the state vectors
of the `parallel' universes which all exist simultaneously, but every of them is characterized by the proper source of the gravitational field
in the particular $k$-th state. Assuming that  the set of vectors $|u_{k} \rangle$ is complete, $\sum_{k} |u_{k} \rangle  \langle u_{k}| = 1$,
we can write
\begin{equation}\label{16}
|\psi \rangle = \sum_{k} |u_{k} \rangle |f_{k} \rangle.
\end{equation}
Substituting Eq. (\ref{16}) into Eq. (\ref{1}), multiplying from the left by $\langle u_{k'}|$, and integrating with respect to the variable
$x$, we obtain the equation for the coefficients $\langle a | f_{k} \rangle = \langle u_{k} | \psi \rangle$,
\begin{equation}\label{17}
\left( - \partial_{a}^{2} + a^{2} - 2 a M_{k} (a) - a^{4} \rho_{\Lambda} - E \right) | f_{k} \rangle = 0.
\end{equation}
The coefficient $| f_{k} \rangle$ is the probability amplitude that the multiverse is in the particular state $|u_{k} \rangle$.

According to the well-known rule, the time derivative of the quantum mechanical mean value of a given operator 
is equal to the mean value of the time derivative of this operator \cite{Lan3}. Thus, taking into account Eq. (\ref{3}), we reduce Eq. (\ref{4})
to the form
\begin{equation}\label{18}
\langle \psi |\left(- \frac{d^{2}a}{dT^{2}} - a + \hat{H}_{\phi} - 3 \hat{L}_{\phi} + 2 a^{3} \rho_{\Lambda}\right)| \psi \rangle = 0.
\end{equation}
After integrating with respect to $x$, using the expansion (\ref{16}), we obtain
\begin{equation}\label{19}
\langle f_{k} |\left(- \frac{d^{2}a}{dT^{2}} - a + M_{k}(a) + 2 a^{3} \rho_{\Lambda}\right)| f_{k} \rangle
- 3 \sum_{k'} \langle f_{k} | \langle u_{k} | \hat{L}_{\phi}| u_{k'} \rangle |f_{k'} \rangle = 0.
\end{equation}

We introduce the energy density of classical source ($\phi$-substance) of the gravitational field as follows
\begin{equation}\label{20}
\rho_{m} = \frac{2 M_{k}(a)}{a^{3}} = 
2 \epsilon_{k} \left(\frac{\lambda_{\alpha}}{2} \right)^{\frac{2}{2 + \alpha}} a^{-\frac{6 \alpha}{2 + \alpha}}.
\end{equation}
The quantity $M_{k}(a) = \frac{1}{2} a^{3} \rho_{m}$ is the proper energy of the $\phi$-substance in the comoving volume 
$\frac{1}{2} a^{3}$. The pressure $p_{m}$ of this $\phi$-substance is determined as the derivative of the proper energy 
with respect to volume taken with the minus sign \cite{Lan2},
\begin{equation}\label{21}
p_{m} = - \frac{2}{3 a^{2}} \frac{d M_{k}(a)}{da}.
\end{equation}
On the other hand, using Eqs. (\ref{2}), (\ref{6}), and (\ref{9}), we find that the density $\rho_{m}$ is the mean value of
the density operator $\hat{\rho}_{\phi}$,
\begin{equation}\label{22}
\rho_{m} = \langle u_{k} | \hat{\rho}_{\phi}| u_{k} \rangle.
\end{equation}
Analogously, the pressure of such a substance can be defined as the mean value of the pressure operator $\hat{p}_{\phi}$,
\begin{equation}\label{23}
p_{m} = \langle u_{k} | \hat{p}_{\phi}| u_{k} \rangle.
\end{equation}
As a result, Eqs. (\ref{17}) and (\ref{19}) take the form
\begin{equation}\label{24}
\langle f_{k} | \left( - \partial_{a}^{2} + a^{2}  - a^{4} \rho \right) | f_{k} \rangle = 0,
\end{equation}
\begin{equation}\label{25}
\langle f_{k} |\left(- \frac{d^{2}a}{dT^{2}} - a + \frac{a^{3}}{2} (\rho - 3p)\right)| f_{k} \rangle
- \frac{3}{2} \langle f_{k} |a^{3} \sum_{k' \neq k} \langle u_{k} | \hat{p}_{\phi} | u_{k'} \rangle |f_{k'} \rangle = 0,
\end{equation}
where we denote
\begin{equation}\label{26}
\rho = \rho_{m} + \rho_{\gamma} + \rho_{\Lambda}, \quad p = p_{m} + p_{\gamma} + p_{\Lambda},
\end{equation}
and take into account the explicit forms of the operator $\hat{L}_{\phi}$ in the variables $a$ and $x$ (see remark after Eq. (\ref{9}))
and the energy density of radiation $\rho_{\gamma} = E / a^{4}$, and the equations of state $p_{\gamma} = \frac{1}{3} \rho_{\gamma}$,
$p_{\Lambda} = - \rho_{\Lambda}$.

Having in mind the future passage to a classical limit, without loss of generality we choose the function $\langle a | f_{k} \rangle$ in the form
\begin{equation}\label{27}
\langle a | f_{k} \rangle = \frac{C_{k}}{\sqrt{\partial_{a} S_{k} (a)}}\ e^{i S_{k}(a)},
\end{equation}
where $C_{k}$ is the constant and, generally speaking, the phase $S_{k}$ can be complex. 
Substituting it into Eqs. (\ref{24}) and (\ref{25}) we arrive at the equations
\begin{equation}\label{28}
(\partial_{a} S_{k})^{2} + a^{2}  - a^{4} \rho = \frac{3}{4} \left(\frac{\partial_{a}^{2} S_{k}}{\partial_{a} S_{k}} \right)^{2} 
- \frac{1}{2} \frac{\partial_{a}^{3} S_{k}}{\partial_{a} S_{k}},
\end{equation}
\begin{equation}\label{29}
\frac{d^{2}a}{dT^{2}} + a - \frac{a^{3}}{2} [\rho - 3(p + p_{Q})] = 0,
\end{equation}
where 
\begin{equation}\label{30}
p_{Q} = \sum_{k' \neq k} \langle u_{k} | \hat{p}_{\phi} | u_{k'} \rangle \frac{C_{k'}}{C_{k}}
\sqrt{\frac{\partial_{a} S_{k}}{\partial_{a} S_{k'}}}\ e^{i(S_{k'} - S_{k})}
\end{equation}
is a quantum correction to the pressure of the $\phi$-substance stipulated by the fact that the wave function $|\psi \rangle$ (\ref{16}) is the superposition
of all possible states of the classical source of the gravitational field $M_{k}(a)$. Within the framework of the multiverse concept, the pressure
(\ref{30}) describes the influence of the `parallel' universes in the states $k' \neq k$ on the universe under consideration (singled out from the rest) 
in the $k$-th state. In our approach, $p_{Q}$ is the only quantity which takes into account that the probability amplitudes (\ref{27}) mix. 
Let us note that the general solution of Eq. (\ref{24}) will have the form of the superposition of the state (\ref{27}) and its conjugate one.

Eqs. (\ref{28}) and (\ref{29}) are exact. They reduce to the equations in the Einstein-Friedmann form, but with quantum correction terms
to the energy density $\rho$ and pressure $p$.

Eqs. (\ref{3}),  (\ref{16}), and (\ref{27}) determine the relation between the phase $S_{k}$ and the momentum $\pi_{a} = - da/ dT$,
\begin{equation}\label{31}
\partial_{a} S_{k} + \frac{i}{2} \frac{\partial_{a}^{2} S_{k}}{\partial_{a} S_{k}} = - \frac{da}{dT}.
\end{equation}
Then Eq. (\ref{28}) takes the form
\begin{equation}\label{32}
\left(\frac{d a}{d T} \right)^{2} + a^{2}  - a^{4} \rho - i \partial_{a}^{2} S_{k}
- \frac{1}{2} \left[\left(\frac{\partial_{a}^{2} S_{k}}{\partial_{a} S_{k}} \right)^{2} 
- \frac{\partial_{a}^{3} S_{k}}{\partial_{a} S_{k}}\right] = 0.
\end{equation}
Passing in Eqs. (\ref{29}) and (\ref{32}) to the proper time $\tau$ and using a dot to denote the time derivative, we obtain
\begin{equation}\label{33}
\left(\frac{\dot{a}}{a} \right)^{2} + \frac{1}{a^{2}} = \rho + \rho_{Q},
\end{equation}
\begin{equation}\label{34}
\left(\frac{\dot{a}}{a} \right)^{2} + \frac{1}{a^{2}} = - \frac{\ddot{a}}{a} + \frac{1}{2} [\rho - 3(p + p_{Q})],
\end{equation}
where
\begin{equation}\label{35}
\rho_{Q} = \frac{2 M_{Q}(a)}{a^{3}}
\end{equation}
is the correction from the quantum source of the gravitational field
\begin{equation}\label{36}
M_{Q}(a) = \frac{Q_{k}(a)}{2 a}
\end{equation}
to the energy density of matter,
where
\begin{equation}\label{37}
Q_{k}(a) = - \partial_{a}^{2} S_{E} + \frac{1}{2} \left[\left(\frac{\partial_{a}^{2} S_{E}}{\partial_{a} S_{E}} \right)^{2} 
- \frac{\partial_{a}^{3} S_{E}}{\partial_{a} S_{E}}\right].
\end{equation}
Here $S_{E} = - i S_{k}$ is the Euclidean phase (the index $k$ is omitted). The quantity $M_{Q}(a) = \frac{1}{2} a^{3} \rho_{Q}$ 
is the proper energy of the quantum source. The extra multiplier $1/a$ in Eq. (\ref{36}) exhibits the relativistic nature of this source.
The pressure produced by it
\begin{equation}\label{38}
P_{Q} = - \frac{2}{3 a^{2}} \frac{d M_{Q}(a)}{da} \equiv p_{Q} + p_{Q\gamma}
\end{equation}
is the sum of the pressures
\begin{equation}\label{39}
p_{Q} = - \frac{1}{3 a^{3}} \frac{d Q_{k}(a)}{da}, \quad p_{Q\gamma} = \frac{1}{3} \rho_{Q},
\end{equation}
where the pressure $p_{Q}$ is the quantum correction (\ref{30}), and $p_{Q\gamma}$ is a correction for relativity. Substituting Eq. (\ref{33}) into (\ref{34}), we obtain the equation
\begin{equation}\label{40}
\frac{\ddot{a}}{a} = - \frac{1}{2} [\rho + \rho_{Q} + 3(p + P_{Q})].
\end{equation}
Differentiating Eq. (\ref{33}) with respect to time $\tau$ and using Eq. (\ref{40}), we find the local law of the energy conservation in the expanding universe in the $k$-th state
\begin{equation}\label{41}
\dot{\rho} + 3 \frac{\dot{a}}{a} (\rho + p) = - \dot{\rho}_{Q} - 3 \frac{\dot{a}}{a} (\rho_{Q} + P_{Q}).
\end{equation}
It describes the transfer of energy between the ordinary matter with the density $\rho$ and the effective matter with the density $\rho_{Q}$
represented by the quantum corrections. The substitution of the explicit expressions for the densities and the pressures from Eq. (\ref{26}), on the one hand, 
and the quantum corrections (\ref{35}) and (\ref{38}), on the other hand, leads to identity.

The two equations from the three (\ref{33}), (\ref{40}), and (\ref{41}) determine the dynamics of the quantum universe defined by the
stress-energy tensor of the perfect fluid which has classical and quantum components with the energy densities
$\rho$ and $\rho_{Q}$, and the pressures $p$ and $P_{Q}$, respectively.

Eqs. (\ref{33}) - (\ref{40}) rewritten for dimensional physical units (see Appendix) demonstrate that quantum corrections 
to the density $\rho$ and pressure $p$ are proportional to $\hbar$. The function $Q_{k}(a)$, 
remaining dimensionless, contains the term 
with the higher derivatives of the phase $S_{E}$ proportional to $\hbar$. Therefore quantum corrections make contributions $\sim \hbar$ and $\hbar^{2}$ 
to the dynamics of the expanding universe. The part of the pressure $p_{Q\gamma}$ related to the relativity of the quantum correction $\rho_{Q}$ 
is proportional to $\hbar c$, as well as the energy density of the ordinary relativistic matter $\rho_{\gamma}$, and, in principle, 
the first cannot be distinguished from the latter. According to (\ref{31}), the Hubble expansion rate $H = \frac{1}{a^{2}} \frac{da}{dT} = \frac{\dot{a}}{a}$ 
contains the term with the higher derivatives of the phase $S_{k}(a)$ which is proportional to $\hbar$. The basic equations of the quantum theory
written in the form of the Einstein equations are expressed via real-valued functions. This leads with necessity to the higher derivatives 
of the Euclidean phase $S_{E}(a)$. This phase can be calculated by considering the dynamics of the universe in imaginary time, i.e. in so-called 
Euclidean part of space-time continuum \cite{Kuz} or directly from Eq. (\ref{28}) (see below).

\section{The properties of matter in the $\phi^{\alpha}$-model}
In the model of interaction (\ref{7}), from Eqs. (\ref{15}), (\ref{20}), and (\ref{21}) for classical source of the gravitational field, it follows that
after averaging with respect to appropriate quantum states (\ref{22}) and (\ref{23}), the scalar field turns into the effective barotopic fluid with the 
equation of state
\begin{equation}\label{42}
p_{m} = \left(\frac{\alpha - 2}{\alpha + 2} \right) \rho_{m}.
\end{equation}
This equation reproduces known equations of state of matter \cite{Kol, Dym} with the only difference that they should be assigned to the particular quantum 
$k$-th state of the scalar field. 

Let us consider a few cases of definite values of $\alpha$ in (\ref{42}).
From Eqs. (\ref{20}) and (\ref{42}), for $\alpha = 0$ we have
\begin{equation}\label{43}
p_{m} = - \rho_{m} \quad \mbox{with} \quad \rho_{m} = \lambda_{0} \epsilon_{k}.
\end{equation}
This equation describes the vacuum (dark energy) in the $k$-th state $|u_{k} \rangle = e^{ikx}$, where $k = \pm \sqrt{\epsilon_{k} - 1}$. 
Eq. (\ref{42}) for $\alpha = 1$ corresponds to the strings in the $k$-th state,
\begin{equation}\label{44}
p_{m} = - \frac{1}{3} \rho_{m} \quad \mbox{with} \quad 
\rho_{m} = \left(\frac{\lambda_{1}}{2} \right)^{2/3} \frac{2 \epsilon_{k}}{a^{2}},
\end{equation}
where the constant $\epsilon_{k}$ takes arbitrary values, $\epsilon_{k} \gtrless 0$, and $|u_{k} \rangle$ is Airy function. Matter in the form of dust is reproduced at $\alpha = 2$,
\begin{equation}\label{45}
p_{m} = 0 \quad \mbox{with} \quad 
\rho_{m} = \left(\frac{\lambda_{2}}{2} \right)^{1/2} \frac{2 \epsilon_{k}}{a^{3}},
\end{equation}
where $\epsilon_{k} = 2 k + 1$, and $k = 0,1,2, \dots$ is a number of non-interacting identical particles with masses $\sqrt{2 \lambda_{2}}$ 
in the state with the wave function $|u_{k} \rangle$ of quantum oscillator \cite{Kuz}. The relativistic matter is described by Eq. (\ref{42}) for $\alpha = 4$,
\begin{equation}\label{46}
p_{m} = \frac{1}{3} \rho_{m} \quad \mbox{with} \quad 
\rho_{m} = \left(\frac{\lambda_{4}}{2} \right)^{1/3} \frac{2 \epsilon_{k}}{a^{4}}.
\end{equation}
This matter component is in the state $|u_{k} \rangle \sim \sqrt{x} Z_{1/6}(i x^{3}/3)$ (see Sect.~3) and
$\epsilon_{k} < \infty$. Eq. (\ref{42}) for $\alpha = \infty$ corresponds to the model with extremely strong self-action of the field $\phi$,
\begin{equation}\label{47}
p_{m} = \rho_{m} \quad \mbox{with} \quad 
\rho_{m} = \frac{2 \epsilon_{k}}{a^{6}}.
\end{equation}
This equation describes the stiff Zel'dovich matter. The wave function $|u_{k} \rangle$ of such a matter has a form (\ref{13}). 
As it was noted in Sect.~3, the equation of state (\ref{47}) may be realized in the very early universe, namely on sub-Planck scales. 
If it was the case in that epoch, while the modern state of matter in our universe is a mixture of $\phi^{0}$ (dark energy) and $\phi^{2}$ (dust) states \cite{Ben, Pla}, 
then one can make a conjecture that the index $\alpha$ in the $\phi^{\alpha}$-model of interaction bears the information about
time, so that $\alpha$ decreases with an increase of $\tau$. The foregoing can be summarized in the table.
\begin{center}
\begin{tabular}{l c l} \hline
Epoch & Interaction & Comments \\ \hline 
sub-Planck & $\phi^{\infty}$ & vanishing coupling constant \\ 
radiation & $\phi^{4}$ & before recombination \\ 
dust & $\phi^{2}$ & non-relativistic matter \\ 
strings & $\phi^{1}$ & cosmological cellular structure \\ 
vacuum & $\phi^{0}$ & accelerating expansion \\ \hline
\end{tabular}
\end{center}
Such a scheme represents correctly the modern views on a whole sequence of changes (transitions from one form to another) 
of the dominating matter component in the universe during its evolution, except the epoch of strings which also should be observed, 
if the modern state of matter in our universe is in fact a mixture of three states $\phi^{2}$, $\phi^{1}$, and $\phi^{0}$.
The observations on distances $R \gtrsim 3 \times 10^{17} R_{Earth}$ reveal the cellular structure \cite{Zel, Got, Ein}. Galaxies, groups and clusters of galaxies are distributed along the chains which form the borders of cells filled with voids. 
The length of the filaments which connect such a network of galaxies, groups and clusters of galaxies are greater than their width and 
much greater than the thickness. Therefore such structures may be interpreted as $\phi^{1}$-states of matter.

From the table given above, it follows that if one does not take quantum corrections into account, matter in the universe will tend to the 
vacuum $\phi^{0}$-state 
approaching the point of the infinite future $\tau \rightarrow \infty$, so that the expansion of the universe will be accelerating.

\section{The quantum source of the gravitational field}
Let us consider the influence of the quantum source (\ref{36}) on the evolution of the universe.
If one neglects the quantum correction $p_{Q}$ (\ref{39}), then in such an approximation $Q_{k}$ is constant, while Eqs. (\ref{33}) and (\ref{40})
contain only quantum corrections $\rho_{Q} = Q_{k} / a^{4}$ and $P_{Q} = \frac{1}{3} \rho_{Q}$ to the energy density and pressure of radiation.
Thus, the $\phi$-substance is characterized by non–zero energy density $\rho_{m}$ (\ref{20}) and pressure $p_{m}$ (\ref{21}).
From Eq. (\ref{30}), it follows that the approximation $p_{Q} = 0$ means the replacement of the wave function $| \psi \rangle$ 
by one term from the sum (\ref{16}). Since both the energy density of radiation $\rho_{\gamma}$ and quantum correction 
$\rho_{Q} \sim 1 / a^{4}$ are proportional to $\hbar c$ (see Sect.~8 and Appendix), the contribution from $\rho_{Q}$ cannot be singled out
against a background of $\rho_{\gamma}$.

The condition $p_{Q} = 0$ is a rough approximation, because it discards the possible quantum effects on the macro scale of the universe.
Let us consider these effects on a few examples.
Using Eq. (\ref{2}) the operator (\ref{5}) can be represented in the form
\begin{equation}\label{48}
\hat{L}_{\phi} = - \frac{1}{2} a^{3} \left(\hat{\rho}_{\phi} + \frac{4}{a^{6}}\partial_{\phi}^{2} \right).
\end{equation}
Then the corresponding term in Eq. (\ref{19}) reduces to
\begin{equation}\label{49}
\sum_{k'} \langle f_{k} | \langle u_{k} | \hat{L}_{\phi}| u_{k'} \rangle |f_{k'} \rangle = 
-\langle f_{k} | M_{k} | f_{k} \rangle + 
\sum_{k'} \langle f_{k} | a^{3} \langle u_{k} | -\frac{2}{a^{6}} \partial_{\phi}^{2} | u_{k'} \rangle |f_{k'} \rangle.
\end{equation}
Here the second term is the state-summed combination of matrix elements of the operator of the doubled kinetic energy of the field $\phi$
between the states $\langle u_{k} |$ and $|u_{k'} \rangle$. 
If the energy which corresponds to this combination compensates the first mass-energy term (the mass of the $\phi$-substance averaged 
over the states $| f_{k} \rangle$), the left-hand side of Eq. (\ref{49}) vanishes and Eq. (\ref{29}) takes the form
\begin{equation}\label{50}
\frac{d^{2}a}{dT^{2}} + a - \frac{a^{3}}{2} [\rho_{m} + 4 \rho_{\Lambda}] = 0.
\end{equation}
Since $4 \rho_{\Lambda} = \rho_{\Lambda} - 3 p_{\Lambda}$, while $\rho_{\gamma} - 3 p_{\gamma} = 0$, then this equation describes
the universe with pressure-free $\phi$-substance (dust with $p_{m} = 0$), radiation and cosmological constant. The second equation (\ref{32})
still contains the quantum correction in the form of $Q_{k}$-term (\ref{37}) proportional to $\hbar$ (see Appendix).

Let us consider another case, when the second term in (\ref{49}) containing the sum over $k'$ vanishes. In this case, the $\phi$-substance
can be considered as a condensate of particles (quanta of the field $\phi$) with zero kinetic energy averaged over all states. 
Eq. (\ref{29}) has a form
\begin{equation}\label{51}
\frac{d^{2}a}{dT^{2}} + a - \frac{a^{3}}{2} [\rho_{m} + 4 \rho_{\Lambda}] + 3 \frac{a^{3}}{2} \rho_{m} = 0,
\end{equation}
where the last term is singled out and written separately, since there are two opportunities to interpret it.

At first, we associate this term with the action of the pressure $p_{m}$ of a condensate. Then Eq. (\ref{51}) transforms into
\begin{equation}\label{52}
\frac{d^{2}a}{dT^{2}} + a - \frac{a^{3}}{2} [\rho - 3 p] = 0,
\end{equation}
where the energy density $\rho$ and pressure $p$ are determined according to (\ref{26}), but with the equation of state of a condensate
\begin{equation}\label{53}
p_{m} = - \rho_{m},
\end{equation}
which demonstrates that, in the approximation under consideration, the $\phi$-substance will be antigravitating matter. 
It, in principle, can play the role of dark energy which has a dynamic quantum nature.

Another opportunity is to assume that the $\phi$-substance is the pressure-free matter ($p_{m} = 0$). In this case, 
the quantum correction in Eq. (\ref{51}) can be interpreted as dark matter with the energy density $\rho_{dm} = 3 \rho_{m}$.
From such a viewpoint, the mass of this `dark matter' exceeds by a factor of 3 that of the non-relativistic $\phi$-substance.
The total energy density $\rho_{m} + \rho_{dm} = 4 \rho_{m}$ is 4 times greater than the energy density (\ref{22}) obtained after averaging of
the Hamiltonian (\ref{2}) over the states of the scalar field. The additional source of the gravitational field with the energy density $\rho_{dm}$
arises as a result of the summation of the matrix elements of the operator $\hat{L}_{\phi}$ (\ref{5}), when the averaged kinetic energy term 
is neglected. This additional condition provides zero pressure of dark matter.

According to (\ref{51}), in the approximation, when the second term in (\ref{49}) containing the sum over $k'$ vanishes,
the second equation should have the form
\begin{equation}\label{54}
\left(\frac{d a}{d T} \right)^{2} + a^{2}  - a^{4} (\rho + \rho_{dm}) = 0.
\end{equation}
Comparing Eq. (\ref{54}) with Eq. (\ref{33}), we find the relations between $\rho_{dm}$ and the quantum correction $\rho_{Q}$, and
between the masses $M_{Q}$ and $M_{k}$,
\begin{equation}\label{55}
\rho_{Q} = \rho_{dm} = 3 \rho_{m}, \quad M_{Q} = 3 M_{k} = const.
\end{equation}
From Eq. (\ref{36}), it follows that 
\begin{equation}\label{56}
Q_{k} = 6 M_{k} a, \quad \mbox{and} \quad P_{Q} = 0.
\end{equation}
Using Eq. (\ref{56}), one can restore the Euclidean phase $S_{E}$ from  Eq. (\ref{37}). Neglecting the term proportional to $\hbar$
(see Appendix) gives
\begin{equation}\label{57}
S_{E} \simeq - M_{k} a^{3}.
\end{equation}
Then, the wave function $|f_{k} \rangle$ has the form
\begin{equation}\label{58}
|f_{k} \rangle \simeq \frac{N_{k}}{a} \exp \left(- M_{k} a^{3} \right),
\end{equation}
where $N_{k}$ is the normalization constant which in the approximation under consideration is to be determined by integrating over 
the domain from the Planck length to infinite $a$.

Matter in the universe mainly consists of antigravitating dark energy and gravitating dark matter. It follows from the analysis above
that one cannot exclude that if not the whole matter-energy, at least a part of its constituents, has a quantum origin. Dark energy as
a condensate and dark matter in the form of the $\phi$-substance appear here as evidence of the quantum nature of the universe
manifesting itself on macroscopic scales.

\section{The exactly solvable $\phi^{2}$-model}
It is of interest to consider the influence of the quantum source (\ref{36}) on the dynamic properties of the universe in the model which 
allows to obtain the exact analytical solution of Eq. (\ref{28}) for the phase $S_{k}(a)$ and to calculate on its basis the function $Q_{k}(a)$ (\ref{37}).
The simplest model is the $\phi^{2}$-model with vanishing cosmological constant, $\Lambda = 0$, when the field $\phi$ oscillates near the point of its true vacuum.
The $\phi$-substance is in the form of dust with the equation of state and the energy density (\ref{45}). It is an aggregate of $k$ identical particles
with the total mass $M = \sqrt{2 \lambda_{2}} (k + \frac{1}{2})$. Then, Eq. (\ref{28}) takes the form
\begin{equation}\label{60}
(\partial_{z}S_{E})^{2} - z^{2} + 2n + 1 = 
- \frac{3}{4} \left(\frac{\partial_{z}^{2}S_{E}}{\partial_{z}S_{E}}\right)^{2}
+ \frac{1}{2} \frac{\partial_{z}^{3}S_{E}}{\partial_{z}S_{E}},
\end{equation}
where we pass to the Euclidean phase $S_{E}$ and introduce the variable $z = a - M$, $M$ is a 
constant\footnote{Throughout this section, the index $k$ is omitted.}.
It was taken into account that Eq. (\ref{17}) with $M_{k}(a) = M$ and $\rho_{\Lambda} = 0$ has the solution, belonging to the discrete spectrum, 
which decreases at infinity for the eigenvalues 
$E = 2n + 1 - M^{2}$, where $n = 0,1,2, \dots$ numbers the states of the universe with given value of mass-energy $M$. 
For the values $z \neq 0$, it is convenient to look for the solution of Eq. (\ref{60}) in the form
\begin{equation}\label{61}
S_{E}(z) = \frac{1}{2} \ln \left[2 \int_{0}^{z}\! dx\, e^{x^{2}} f(x) \right] + const.
\end{equation}
Substituting Eq.~(\ref{61}) into Eq.~(\ref{60}), we obtain the nonlinear equation for the unknown function $f(z)$
\begin{equation}\label{62}
\frac{1}{2} \frac{\partial_{z}^{2} f}{f} - z \frac{\partial_{z} f}{f} 
- \frac{3}{4} \left(\frac{\partial_{z} f}{f}\right)^{2} = 2n.
\end{equation}
By substituting
\begin{equation}\label{63}
f(z) = H_{n}^{-2}(z),
\end{equation}
it reduces to the equations for the Hermitian polynomials $H_{n}(z)$
\begin{equation}\label{64}
\partial_{z}^{2} H_{n} - 2z \partial_{z}H_{n} + 2n H_{n} = 0.
\end{equation}
From Eqs. (\ref{61}) and (\ref{63}), we find the derivative of $S_{E}$ with respect to $z$
\begin{equation}\label{65}
\partial_{z} S_{E} = \frac{e^{z^{2}} H_{n}^{-2}(z)}{2\int_{0}^{z}\! dx\,e^{x^{2}} H_{n}^{-2}(x)}.
\end{equation}
Passing in Eq. (\ref{37}) to the variable $z$, using (\ref{65}) and the properties of the Hermitian polynomials, we obtain
\begin{equation}\label{66}
Q_{n}(a) = -1 - 2n \chi_{n} (a - M),
\end{equation}
where we denote
\begin{equation}\label{67}
\chi_{n}(z) = 1 - \frac{H_{n-1}(z) H_{n+1}(z)}{H_{n}^{2}(z)}.
\end{equation}
The quantum correction $\rho_{Q}$ (\ref{35}) to the energy density $\rho$ (\ref{26}) and the pressure $p_{Q}$ (\ref{39}) are
\begin{equation}\label{68}
\rho_{Q} = - \frac{1}{a^{4}} - \frac{n}{a^{6}}\mu_{n}(z), \quad
p_{Q} = - \frac{2n}{3a^{6}}b_{n}(z),
\end{equation}
where
\begin{equation}\label{69}
\frac{\mu_{n}}{\mu_{1}} = 2 z^{2} \chi_{n}(z), \quad
\frac{b_{n}}{b_{1}} = - z^{3} \frac{d\chi_{n}(z)}{dz},
\end{equation}
and
\begin{equation}\label{70}
\mu_{1} = \left(\frac{a}{z}\right)^{2}, \quad
b_{1} = \left(\frac{a}{z}\right)^{3}.
\end{equation}
At the point $z = 0$, the functions $\frac{\mu_{n}}{\mu_{1}}$ and $\frac{b_{n}}{b_{1}}$ vanish for even values of $n$ and equal to a constant for odd $n$. 
In the region $|z| \gg 1$, the functions $\mu_{n} \simeq 1$ and $b_{n} \simeq 1$.
We give the explicit forms of these functions for small $n$,
\begin{equation}\label{71}
\frac{\mu_{2}}{\mu_{1}} = \frac{z^{2} (z^{2} + \frac{1}{2})}{(z^{2} - \frac{1}{4})^{2}}, \quad
\frac{\mu_{3}}{\mu_{1}} = \frac{(z^{4} + \frac{3}{4})}{(z^{2} - \frac{3}{2})^{2}},
\quad
\frac{\mu_{4}}{\mu_{1}} = \frac{z^{2}(z^{6} + \frac{3}{2}z^{4} + \frac{9}{4} z^{2} 
+ \frac{9}{8})}{(z^{4} - 3z^{2} + \frac{3}{4})^{2}},
\end{equation}
\begin{equation}\label{72}
\frac{b_{2}}{b_{1}} = \frac{z^{4}}{(z^{2} - \frac{1}{4})^{2}} 
\left[2\, \frac{z^{2} + \frac{1}{2}}{z^{2} - \frac{1}{4}} - 1 \right], \quad
\frac{b_{3}}{b_{1}} = \frac{z^{4}}{\left(z^{2} - \frac{3}{2}\right)^{2}} 
\left[3\, \frac{(z^{4} + \frac{3}{4}) (z^{2} - \frac{1}{2})}{z^{4} (z^{2} - \frac{3}{2})} - 2 \right],
\end{equation}
and so on.

Using the properties of the Hermitian polynomials, for both asymptotic cases $|z| \rightarrow \infty$, $n < \infty$ and $n \rightarrow \infty$, $|z| < \infty$, we find that $Q_{n}(a) \sim -1$ and 
\begin{equation}\label{73}
\rho_{Q} \sim - \frac{1}{a^{4}}, \quad p_{Q} \sim 0.
\end{equation}
The mass-energy of the quantum source turns out to be negative
\begin{equation}\label{74}
M_{Q}(a) \sim - \frac{1}{2 a}.
\end{equation}
In the state $n = 0$, the relations (\ref{73}) and (\ref{74}) become exact equalities.

\section{The Weyssenhoff fluid}
Using the explicit form of $\rho_{Q}$ and $p_{Q}$ (\ref{68}), we rewrite Eqs. (\ref{33}) and (\ref{40}) for $\rho_{\Lambda} = 0$ as follows
\begin{equation}\label{75}
\left(\frac{\dot{a}}{a} \right)^{2} + \frac{1}{a^{2}} = \rho_{m} + \frac{E - 1}{a^{4}} - \frac{n}{a^{6}}\, \mu_{n},
\end{equation}
\begin{equation}\label{76}
\frac{\ddot{a}}{a} = - \frac{1}{2} \left[\rho_{m} + 3 p_{m} + 2 \frac{E - 1}{a^{4}} - \frac{2 n}{a^{6}} (\mu_{n} + b_{n})\right].
\end{equation}
In order to establish the physical meaning of the terms with $a^{-6}$, we convert these equations to ordinary units taking $a$ and $\tau$ in cm,
$\rho_{m}$ and $p_{m}$ in GeV/cm$^{3}$, and leaving $E$, $\mu_{n}$, and $b_{n}$ dimensionless,
\begin{equation}\label{77}
\left(\frac{\dot{a}}{a} \right)^{2} + \frac{1}{a^{2}} = \frac{8 \pi G}{3 c^{4}} \left[\rho_{m} + \frac{\hbar c}{4 \pi^{2}} \frac{E - 1}{a^{4}} 
- \frac{2 \pi G}{c^{2}} \sigma^{2}\right],
\end{equation}
\begin{equation}\label{78}
\frac{\ddot{a}}{a} = - \frac{4 \pi G}{3 c^{4}} \left[\rho_{m} + 3 p_{m} + \frac{\hbar c}{2 \pi^{2}} \frac{E - 1}{a^{4}} 
- \frac{4 \pi G}{c^{2}} \sigma'^{2} \right],
\end{equation}
where
\begin{equation}\label{79}
\sigma^{2} = \frac{\hbar^{2}}{12 \pi^{4}}\frac{n \mu_{n}}{a^{6}}, \quad
\sigma'^{2} = \sigma^{2} + \frac{\hbar^{2}}{12 \pi^{4}}\frac{n b_{n}}{a^{6}}.
\end{equation}
Up to quantum correction which contributes to $\sigma'^{2}$, Eqs. (\ref{77}) and (\ref{78}) may be recognized as the equations of the 
Einstein-Cartan theory of gravity with torsion for the FRW universe \cite{Kop,Nur,Gas}.
According to Cartan, the antisymmetric part of the affine connection coefficients (torsion) becomes an independent dynamic variable which can
be associated with the spin density of matter in the universe \cite{Heh}.
Eqs. (\ref{77}) and (\ref{78}) can be considered as describing the homogeneous, isotropic and spatially closed universe filled with the
$\phi$-substance in the form of a perfect fluid with spin. Such a fluid, often called Weyssenhoff fluid \cite{Wey}, is a perfect incompressible fluid 
(continuous medium) every element of which is interpreted as a particle with spin. Such a spin fluid is characterized by the energy density
$\rho_{m}$, pressure $p_{m}$ and proper angular momentum density, or spin density $s_{\mu \nu}$. In the Einstein–Cartan theory,
the value $\sigma^{2}$ is the square of the spin density $\sigma^{2} = \frac{1}{2} \langle s_{\mu \nu}  s^{\mu \nu} \rangle$, where
the suitable space-time averaging is performed in order to make a transition to macroscopic scales. It is assumed that the spins are not polarized,
but are randomly oriented, so that the average $\langle s_{\mu \nu}\rangle = 0$.

If the spin fluid consists of baryons with spin $\frac{\hbar}{2}$, the average total spin density is equal to $s = \frac{\hbar}{2} \mbox{n}$, 
where $\mbox{n} = \frac{N}{V}$ is the average baryon number density, $N$ is the number of baryons contained in the volume
$V \sim a^{3}$. In this case, the square of the spin density $\sigma^{2}$ has a simple form $\sigma^{2} = \frac{\hbar^{2}}{8} \mbox{n}^{2}$ \cite{Nur},
i.e. $\sigma^{2} \sim \frac{\hbar^{2}}{a^{6}}$. Identifying this quantity with $\sigma^{2}$ of Eq. (\ref{79}), we obtain the expression for
the number of spin particles in the volume $V = 2 \pi^{2} a^{3}$,
\begin{equation}\label{80}
N = \sqrt{\frac{8}{3} n \mu_{n}}.
\end{equation}
Eq. (\ref{17}) for $\rho_{\Lambda} = 0$ has the solution  decreasing at infinity for the values $E - 1 = 2n - M^{2}$, where $M$ is the total mass of 
$\phi$-substance in the universe taken in units of Planck mass $m_{P}$, i.e. the number of particles with masses $m_{P}$. For the epoch
$(E - 1)/ M^{2} \ll 1$, we have $2 n \simeq M^{2}$ and
\begin{equation}\label{81}
N \simeq 2 M \sqrt{\frac{\mu_{n}}{3}}.
\end{equation}
Let us estimate the number $N$ for the observed part of our universe. Setting the energy density of radiation $\rho_{\gamma} \sim 10^{-10}$ GeV /cm$^{3}$,
the quantity of matter $M \sim 10^{57}$ g, the radius $a \sim 10^{28}$ cm, we find that $E \sim 10^{118}$, $M \sim 10^{61}$. Thus,
$(E - 1)/ M^{2} \sim 10^{-4}$ and the formula (\ref{81}) is valid. Supposing that  the main contribution into the mass $M$ is made
by baryons with masses $\sim 1$ GeV (protons) and taking $\mu_{n} \sim O(1)$, we obtain 
that today the equivalent number of baryons in the observed part of the universe is $N \sim 10^{80}$. This estimation coincides completely with estimation of the
number of baryons which follows from the equations of general relativity for closed universe \cite{Mis}, with the only addition that
these baryons are particles with spin $\frac{\hbar}{2}$.
Since $b_{n} \sim \mu_{n}$, the quantum correction $p_{Q}$ (\ref{68}) makes the same contribution into Eq. (\ref{78})
as the correction $\rho_{Q}$ to the energy density, $\sigma'^{2} \sim 2 \sigma^{2}$. The identification of $\sigma^{2}$ (\ref{79})
with the square of the spin density reveals the quantum nature of this characteristic of matter in the universe.

\section{Conclusion}
The study of the dynamics of the expanding universe with regard for possible quantum effects is of interest in view of evidence 
from astronomical data about the matter-energy content in our universe interpreted in favour of the existence of the mysterious dark
matter and dark energy \cite{Ben, Pla}. In the present paper, by the example of the FRW universe with cosmological constant, 
originally filled with a uniform scalar field $\phi$ and radiation, it is shown that the 
equations of the quantum theory can be formally reduced to the exact Einstein-type equations with an additional source 
of the gravitational field of quantum nature. Quantum corrections to the energy density and pressure are proportional to $\hbar$ and $\hbar^{2}$.
After averaging with respect to appropriate quantum states, the scalar field turns into the effective barotopic fluid whose properties
are determined by the form of the primordial scalar field potential.
They are summarized in Table (see Sect.~5). By modifying the contribution from
the kinetic term of the averaged scalar field, both pressure-free matter and matter with the vacuum-type 
equation of state can be reproduced by the effective matter-energy brought about by quantum corrections. 
It is shown that quantum effects into the gravitational 
interaction can be significant on macroscopic scale: unless the whole, at least a part of such 
matter-energy constituents as dark matter and dark energy may have a quantum origin.
Quantum equations of the universe in the pure quantum state in the $\phi^{2}$-model are identical to the
the equations of the Einstein-Cartan theory of gravity with torsion.
After averaging over its quantum states, the free scalar field in the $\phi^{2}$-model turns into the Weyssenhoff fluid characterized
by the energy density, pressure, and spin of constituent particles. 
The correspondence between the equivalent number of baryons (protons) in the observed part of the universe and the
value of spin of these particles being equal to $\hbar^{2}/2$ is found.

\appendix
\section{Appendix}
Measuring the scale factor $a$ and time $\tau$ in cm, the energy density and pressure in GeV/cm$^{3}$, and the phase $S_{k} (a)$
in cm$^{2}$, we reduce Eqs. (\ref{28}) and (\ref{31}) to the form
\begin{equation}\label{82}
(\partial_{a} S_{k})^{2} + a^{2}  - \frac{8 \pi G}{3 c^{4}} a^{4} \rho = 
l_{P}^{4} \left[\frac{3}{4} \left(\frac{\partial_{a}^{2} S_{k}}{\partial_{a} S_{k}} \right)^{2} 
- \frac{1}{2} \frac{\partial_{a}^{3} S_{k}}{\partial_{a} S_{k}}\right],
\end{equation}
\begin{equation}\label{83}
\partial_{a} S_{k} + i \frac{l_{P}^{2}}{2} \frac{\partial_{a}^{2} S_{k}}{\partial_{a} S_{k}} = - \frac{da}{dT}, \quad
 \frac{da}{dT} = a \dot{a},
\end{equation}
where $l_{P} = \sqrt{2 G \hbar / (3 \pi c^{3})}$ is the Planck length and the time variable $T$ is dimensionless,
$G$ is the Newtonian gravitational constant and the ratio $c^{4} / G$ is measured in GeV/cm. 
In this units Eqs. (\ref{29}) and (\ref{32}) are
\begin{equation}\label{84}
\frac{d^{2}a}{dT^{2}} + a - \frac{4 \pi G}{3 c^{4}} a^{3} [\rho - 3(p + p_{Q})] = 0,
\end{equation}
\begin{equation}\label{85}
\left(\frac{d a}{d T} \right)^{2} + a^{2}  - \frac{8 \pi G}{3 c^{4}} a^{4} \rho - l_{P}^{2} Q_{k} (a)= 0,
\end{equation}
where the pressure
\begin{equation}\label{86}
p_{Q} = - \frac{\hbar c}{12 \pi^{2} a^{3}} \frac{d Q_{k}}{da},
\end{equation}
and
\begin{equation}\label{87}
Q_{k}(a) = - \partial_{a}^{2} S_{E} + \frac{l_{P}^{2}}{2} \left[\left(\frac{\partial_{a}^{2} S_{E}}{\partial_{a} S_{E}} \right)^{2} 
- \frac{\partial_{a}^{3} S_{E}}{\partial_{a} S_{E}}\right]
\end{equation}
is the dimensionless function of $a$. It contains the term proportional to $\hbar$. Eqs. (\ref{33}) and (\ref{40}) have the same form as
the equations of general relativity for the FRW universe. Using Eqs. (\ref{35}) and (\ref{36}), they can be written as
\begin{equation}\label{88}
\left(\frac{\dot{a}}{a} \right)^{2} + \frac{1}{a^{2}} = \frac{8 \pi G}{3 c^{4}} \left[\rho_{m} + \frac{\hbar c}{4 \pi^{2}} \frac{E + Q_{k}}{a^{4}} \right] +
\frac{\Lambda}{3},
\end{equation}
\begin{equation}\label{89}
\frac{\ddot{a}}{a} = - \frac{4 \pi G}{3 c^{4}} \left[\rho_{m} + 3p_{m} + \frac{\hbar c}{2 \pi^{2}} \left(\frac{E + Q_{k}}{a^{4}} 
- \frac{1}{2 a^{2}} \frac{d Q_{k}}{da}\right)\right] + \frac{\Lambda}{3},
\end{equation}
where the explicit forms of the energy densities of radiation $\rho_{\gamma}$ and vacuum  $\rho_{\Lambda}$ were taken into account,
\begin{equation}\label{90}
\rho_{\gamma} = \frac{\hbar c}{4 \pi^{2}} \frac{E}{a^{4}}, \quad \rho_{\Lambda} = \frac{\Lambda c^{4}}{8 \pi G}.
\end{equation}
The constant $E$ is dimensionless. Its numerical value is determined from the solution of eigenvalue problem (\ref{17}). 
The cosmological constant is measured in cm$^{-2}$. 

If $Q_{k}$ is a sum of terms one of which does not depend on $a$, this term contributes to the energy density of radiation.

From the given equations, it follows that the quantum corrections make a contribution to the dynamics of the universe proportional to
$\hbar$ and $\hbar^{2}$. But, as it is shown in Sect.~6 and 7, the quantities standing after these constants can under specific conditions 
appear to be very large ($\sim \hbar^{-1}$ and $\hbar^{-2}$), so that the contribution of quantum effects to the gravitational interaction 
can be significant not only on micro, but also on macro scale.

\end{document}